# Accuracy Enhancement in Refractive Index Sensing via Full-Spectrum Machine Learning Modeling

Majid Aalizadeh[1,2,3,4], Chinmay Raut[5], Morteza Azmoudeh Afshar[6], Ali Tabartehfarahani[1,3,4], and Xudong Fan[1,3,4,*]

[1]Department of Biomedical Engineering,
University of Michigan, Ann Arbor, MI 48109, USA

[2]Department of Electrical Engineering and Computer Science,
University of Michigan, Ann Arbor, MI 48109, USA

[3]Center for Wireless Integrated MicroSensing and Systems (WIMS[2]),
University of Michigan, Ann Arbor, MI 48109, USA

[4]Max Harry Weil Institute for Critical Care Research and Innovation,
University of Michigan, Ann Arbor, MI 48109, USA

[5]Department of Computational Medicine and Bioinformatics,
University of Michigan, Ann Arbor, MI 48109, USA

[6]Informatics Institute
Istanbul Technical University, 34485 Istanbul, Turkey

*: Corresponding author: xsfan@umich.edu

**Abstract**

We present a full-spectrum machine learning framework for refractive index sensing using simulated absorption spectra from meta-grating structures composed of titanium or silicon nanorods under TE and TM polarizations. Linear regression was applied to 80 principal components extracted from each spectrum, and model performance was assessed using five-fold cross-validation, simulating real-world biosensing scenarios where unknown patient samples are predicted based on standard calibration data. Titanium-based structures, dominated by broadband intensity changes, yielded the lowest mean squared errors and the highest accuracy improvements—up to a 6065-fold reduction compared to the best single-feature model. In contrast, silicon-based structures, governed by narrow resonances, showed more modest gains due to spectral nonlinearity that limits the effectiveness of global linear models. We also show that even the best single-wavelength predictor is identified through data-driven analysis, not visual selection, highlighting the value of automated feature preselection. These findings demonstrate that spectral shape plays a key role in modeling performance and that full-spectrum linear approaches are especially effective for intensity-modulated index sensors.

## 1. Introduction

Resonant nanophotonic structures are widely used in optical biosensing, where changes in the surrounding refractive index produce measurable shifts in the optical spectrum [1-18]. These shifts are typically tracked through single resonance peaks or simplified one-dimensional fitting routines that monitor only a small portion of the spectrum. While these methods are practical and interpretable, they inherently overlook the rich spectral features that span across broader wavelength ranges and may contain valuable sensing information.

Machine learning has been increasingly used across scientific domains, including biosensing, to uncover patterns in complex data and enhance predictive performance [19-26]. In our prior work[27], we introduced a machine learning approach that used multiple pre-selected resonance peak shifts as inputs for ridge regression. That method significantly reduced mean squared error in refractive index prediction, outperforming conventional single-feature techniques by orders of magnitude. However, it still required manual feature selection and was limited to spectra with well-defined peaks.

In parallel, the advancement of optical sensing hardware has been accelerated by the development of metamaterials. These are artificially structured materials engineered to manipulate electromagnetic waves beyond the capabilities of naturally occurring media[28]. Metasurfaces represent the two-dimensional class of metamaterials and offer powerful control over phase, amplitude, and polarization[29-33]. Among them, meta-gratings are periodic metasurfaces with a wide range of applications and are highly suited for biosensing applications due to their strong spectral sensitivity and tunability[34].

In this work, we explore whether full-spectrum machine learning, without manual feature selection, or peak selection, can improve prediction accuracy across different spectral response types. We simulate a meta-grating structure consisting of triangular nanorods on a gold reflector, where the rod material is set to either Ti (Ti) or Si (Si). Ti supports broadband intensity spectrum modulation, while Si supports narrowband shifting Mie-type resonances. To account for the anisotropy of the design, both TE and TM polarizations are simulated for each configuration.

To evaluate generalization, we apply five-fold cross-validation by holding out different refractive index values during training. This mimics experimental biosensing scenarios, where standard samples form a calibration curve and unknown analytes are predicted. Traditional methods rely on one-dimensional fitting at a visually selected wavelength, but our results show that even the best single predictor is identified through machine learning rather than visual inspection. Full-spectrum modeling with principal component analysis (PCA) and linear regression consistently improves accuracy, with up to a 6000-fold mean squared error (MSE) reduction in Ti-based structures due to their smooth, intensity-driven spectra. These findings underscore the critical role of spectral shape and highlight the advantage of full-spectrum approaches, particularly for meta-gratings governed by broadband intensity modulation.

## 2. Proposed Structure

Figure 1 illustrates the meta-grating structure used as the bulk index sensing platform in this study. The geometry consists of a one-dimensional periodic array of triangular cross-sectioned nanorods positioned on an optically thick gold (Au) reflector. The simulations have revealed that other metals can be used as the back reflector as well, given that they are optically thick to block the transmission. Each nanorod has a base (period) and height of 3 μm, forming an isosceles triangular cross-section. The material of the nanorods is set to either Ti or Si, allowing for the exploration of different resonance behaviors. Ti supports broadband intensity-modulated spectra, while Si enables narrowband Mie-type resonances with distinct peak shifts. This design serves as a unified geometry for comparing different spectral modalities within the same structural platform.

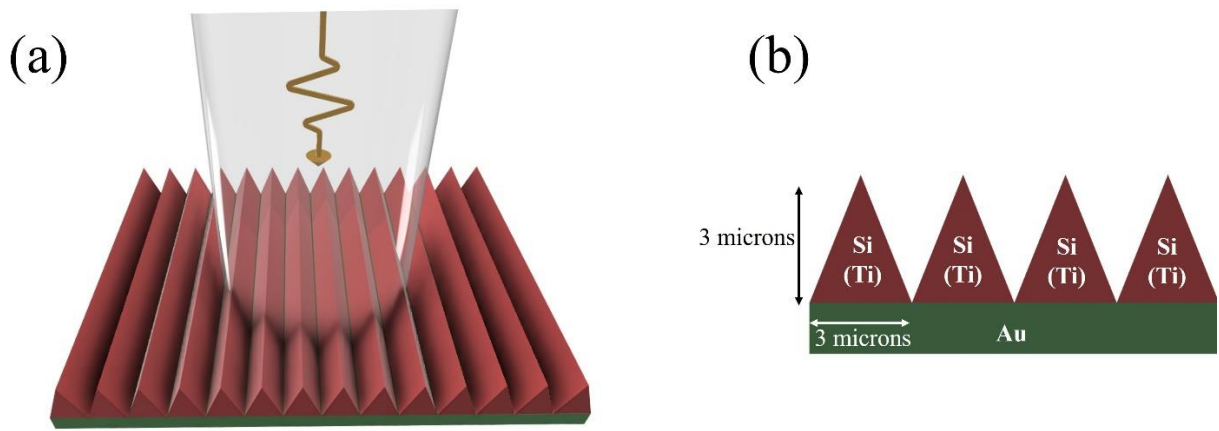

Figure 1. (a) Schematic, and (b) cross section side view of the meta-grating refractive index sensor consisting of triangular periodic nanorods on a gold back reflector. Both Si and Ti materials are used as triangular nanords.

Figure 1(a) presents a perspective view of the meta-grating illuminated under normal incidence, while Figure 1(b) demonstrates the cross-sectional profile of 4 neighboring unit cells. The simulations are carried out by systematically varying the refractive index of the surrounding environment above the meta-grating to evaluate the sensing response. The periodic triangular profile supports multi-resonant behavior and strong light–matter interaction, making it suitable for assessing the performance of full-spectrum machine learning models under varying physical regimes.

## 3. Simulation Results

Figure 2 presents the absorption spectra of the meta-grating for different material and polarization combinations, with the refractive index of the environment fixed at 1. Figures 2(a) and (b) correspond to the Ti-based meta-grating under TM and TE polarizations, respectively. Both configurations display smooth, broadband absorption profiles with low spectral feature density. The spectra lack sharp resonances and instead show gradual intensity variations across the wavelength range. This behavior is characteristic of lossy metallic nanostructures and suggests that the optical response is dominated by broadband absorption rather than discrete resonance effects.

Figures 2(c) and (d) show the Si-based meta-grating under TM and TE polarizations, respectively. In these cases, the spectra exhibit a series of narrow, well-resolved resonances, indicative of Mie-type modes supported by high-index dielectric materials[27]. The spectral profiles show greater feature density and sharper variations compared to the Ti cases, and the position and spacing of the resonances differ slightly between TE and TM modes due to the anisotropic response of the structure. These differences in spectral shape across materials and polarizations set the foundation for evaluating how spectral modality influences modeling strategies in subsequent analyses.

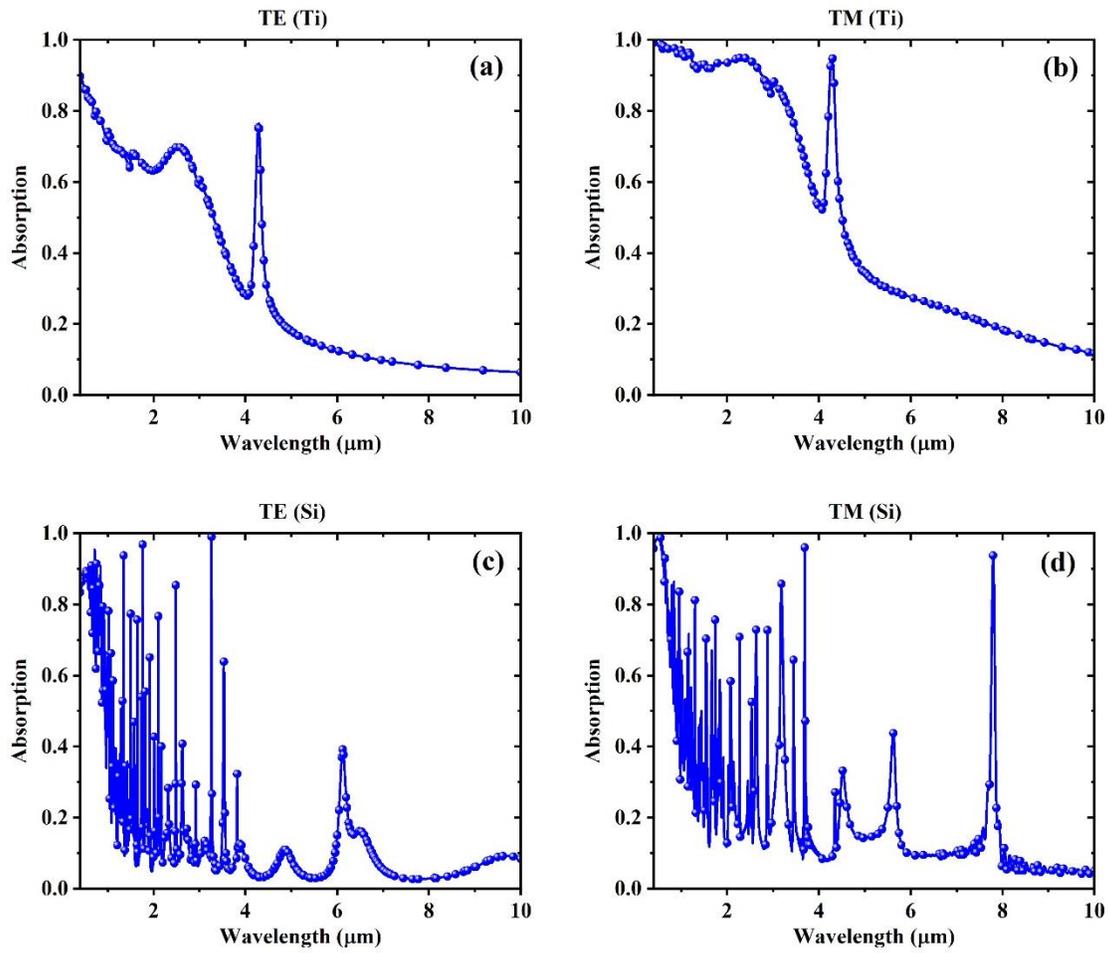

Figure 2. Absorption spectra of the structure at the environment refractive index of 1, using (a) Ti with TM (b) Ti with TE, (c) Si with TM, and (d) Si with TE polarization.

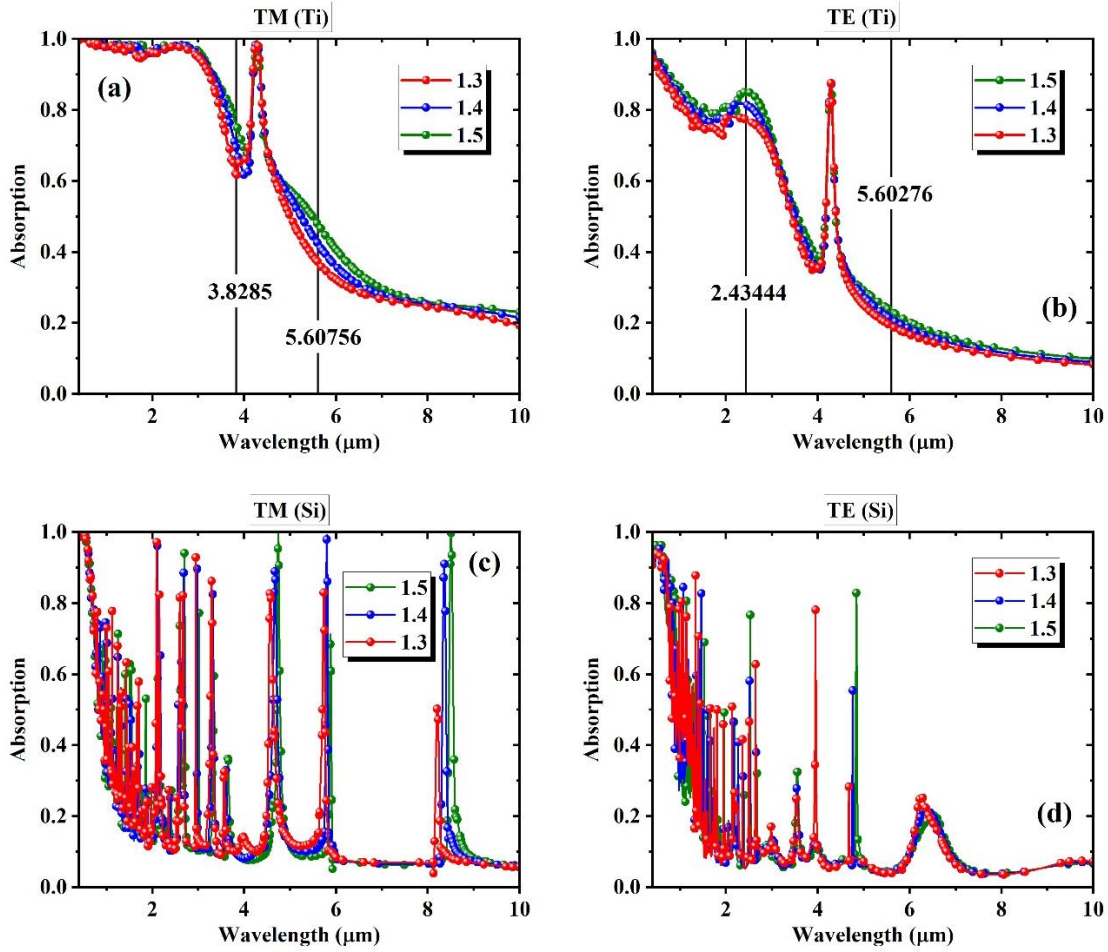

Figure 3. Absorption spectra of the structure at the environment with varying refractive indices of 1.3, 1.4, and 1.5 using (a) Ti with TM, (b) Ti with TE, (c) Si with TM, and (d) Si with TE polarization.

Figure 3 shows how the absorption spectra change with variations in the refractive index of the surrounding environment, ranging from 1.3 to 1.5. The meta-grating is simulated under TM and TE polarizations for both Ti and Si nanorods. Figures 3(a) and (b) correspond to the Ti-based structures, where the overall spectral shape remains consistent across different refractive indices. The dominant change is in the intensity of absorption at specific regions, rather than shifts in spectral position. This is more evident by the only resonance almost not shifting at all by varying the index. A few wavelengths that appear to exhibit relatively strong monotonic intensity changes are marked on the plots. These points represent wavelengths with the most prominent index-dependent intensity changes and are typically chosen in intensity-modulation–based biosensing approaches.

Figures 3(c) and (d) show the Si-based spectra under TM and TE polarizations. Unlike the Ti cases, the spectra here contain dense, sharp resonances that shift in wavelength as the refractive index varies. These features reflect Mie-type behavior typical of high-index dielectric nanostructures. The field analysis representing the physical phenomena underlying such resonances is thoroughly discussed in our prior work[27]. Because the peak shifts are narrow and localized, the intensity at most fixed wavelengths remains relatively constant unless a resonance aligns with that location. This creates a more complex spectral behavior that is less compatible with simple linear modeling. By contrast, the Ti spectra display smoother and more predictable global changes in intensity, which are better suited for linear regression. These observations establish the foundation for comparing full-spectrum machine learning performance across different structural and spectral regimes.

Figure 4 provides magnified views of the selected spectral regions from the Ti-based meta-grating, focusing on wavelengths with high variations of intensity across refractive indices. Figures 4(a–d) show the spectral response at four different wavelengths, chosen based on visually identifiable intensity changes across refractive indices. These wavelengths were selected because they appear to vary in a monotonic and relatively consistent manner, which aligns with the assumptions of linear modeling. Figures 4(a) and (c) show TM polarization results, while Figures 4(b) and (d) correspond to TE polarization. Each of these points is used to assess how well a single-wavelength intensity value can serve as a predictor for refractive index using conventional regression.

Figures 4(e) and (f) present the same Ti-based spectra, zoomed in around the single spectral peak in TM and TE polarizations, respectively. These plots demonstrate that, unlike the Si-based structures, the single peak position does not shift noticeably with changes in refractive index. The spectral shape remains fixed, and the only observable variation is in the amplitude of absorption. This highlights a key distinction in the sensing behavior of the Ti structure. While the intensity-based features allow for straightforward linear modeling, there is no resonant wavelength tracking involved. These observations confirm that intensity modulation is the dominant mechanism for index detection in Ti-based configurations.

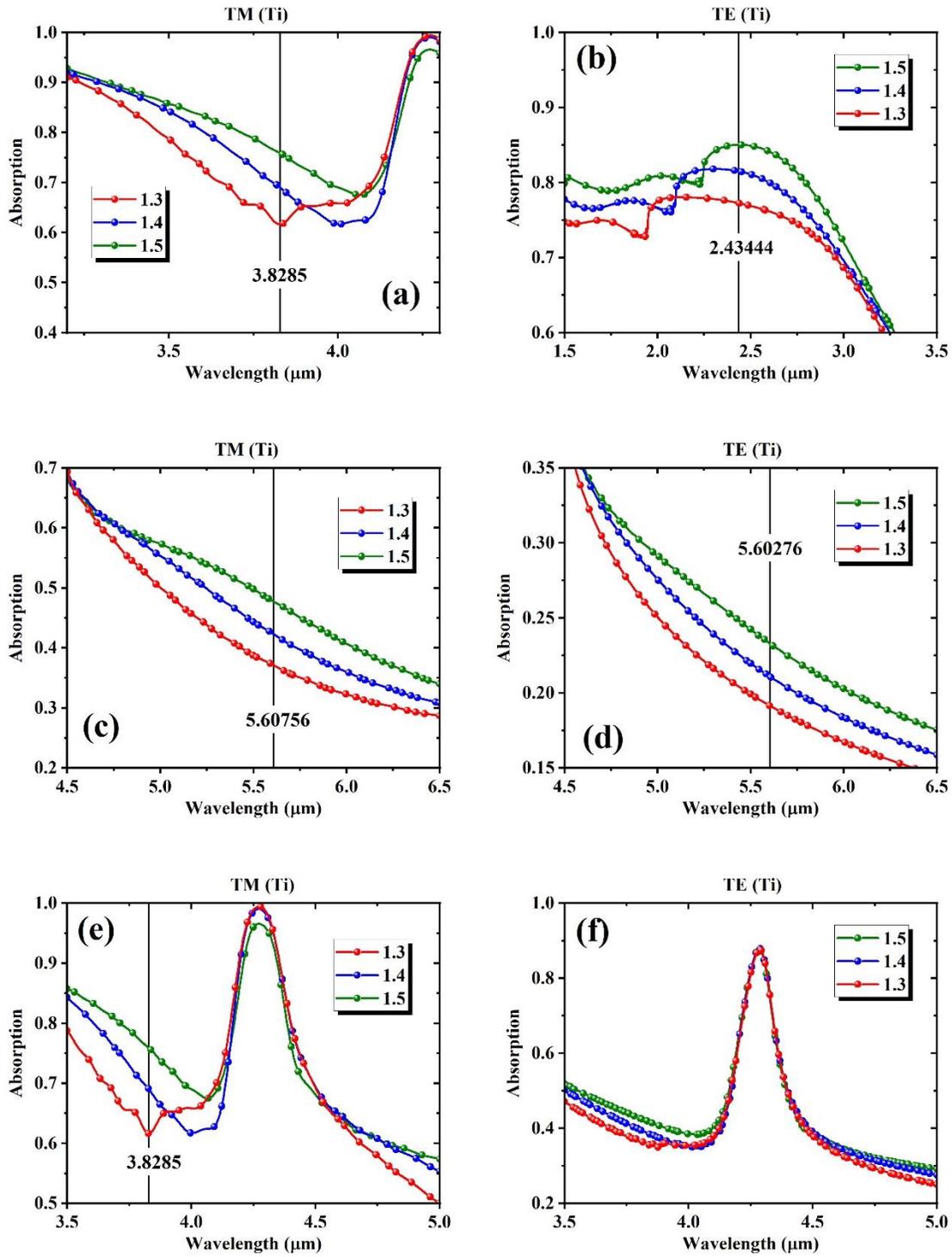

Figure 4. Localized absorption intensity variation in Ti-based structures for TM (left panel) and TE (right panel) polarizations. (e) and (f) demonstrate nearly no peak shift for varying index in the Ti-based structures.

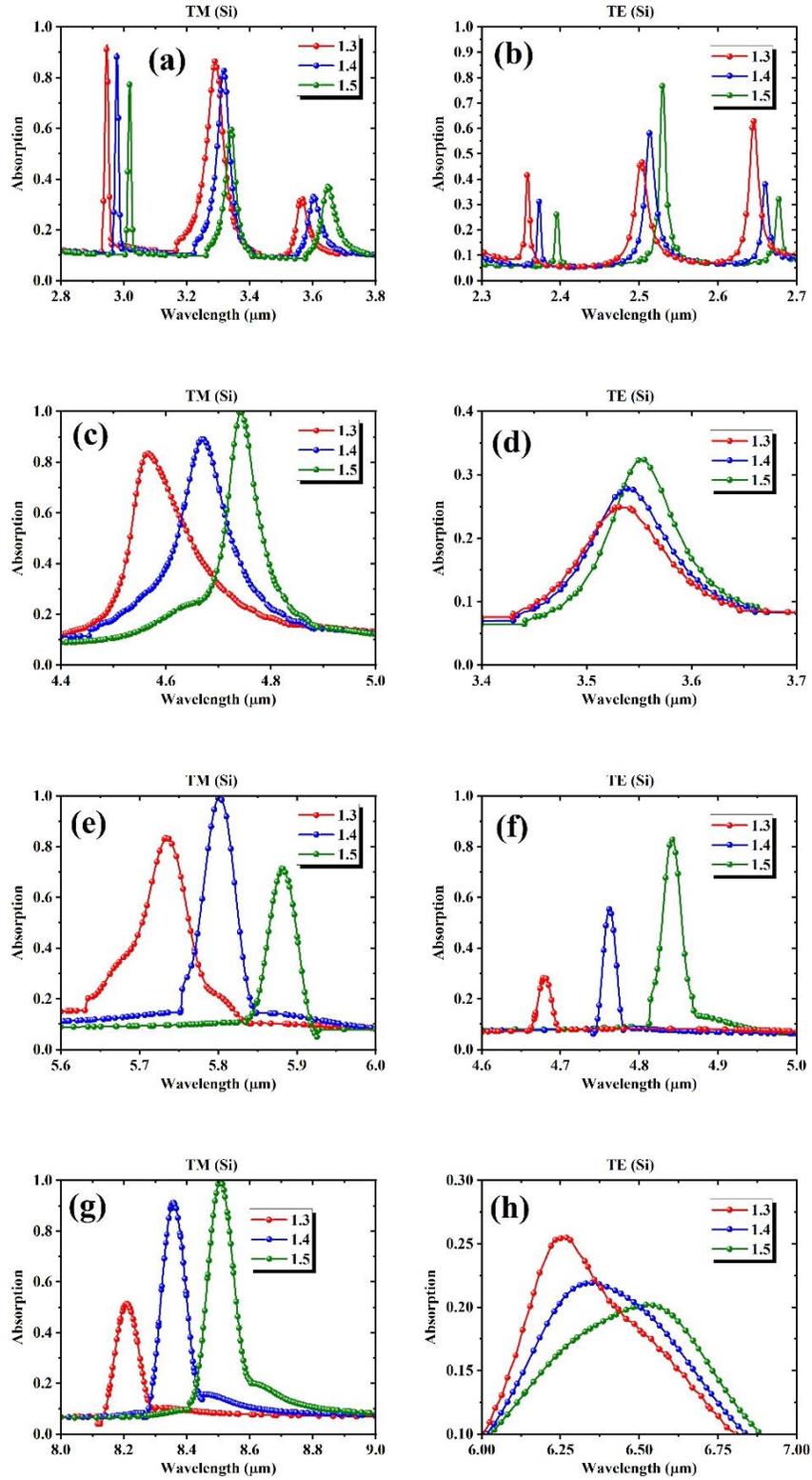

Figure 5. Localized Resonance Shifts in Si-Based Structures for TM (left panel) and TE (right panel) polarizations.

Figure 5 presents a series of zoomed-in absorption spectra for the Si-based meta-grating under TM and TE polarizations, illustrating sharp resonances that shift significantly as the bulk refractive index changes from 1.3 to 1.5. Each subpanel highlights one or more narrow Mie-type resonances and their red-shift across the index increase. The TM mode spectra are shown in the left panel, or Figs. 5(a), (c), (e), and (g), while TE mode results are given in the right panel, or Figs. 5(b), (d), (f), and (h). The selected peaks exhibit high spectral sensitivity and sharpness, with shifts that span tens of nanometers or more for a 0.1 variation in the bulk index. This results in nm/RIU in the order of tens. These shifts often vary in slope and symmetry depending on the peak's location and the surrounding spectral context.

These peak shifts, which occur consistently across a range of wavelengths, form the foundation of traditional biosensing approaches that rely on tracking the movement of a single resonance. Figure 6 further quantifies this behavior by plotting the wavelength shifts of selected resonances in the Si-based meta-grating as a function of bulk refractive index, revealing strong linearity for individual peaks. The TM polarization results are displayed in Figure 6(a), and TE results are shown in Figure 6(b). For both polarizations, the tracked peaks exhibit consistent redshifts as the refractive index increases from 1.30 to 1.40, confirming their suitability for refractive index sensing based on spectral displacement. The shift trends appear approximately linear over the simulated range for most peaks, although the magnitude of sensitivity varies substantially across resonance locations. Notably, the 8209 nm peak in the TM configuration shows both the largest and most linear shift, which, as will be discussed later, results in the lowest mean squared error among all single-predictor models in our regression analysis. This observation reinforces the strong link between linearity of spectral displacement and predictive performance in linear modeling frameworks.

When the peak tracing behavior is similar to the one shown in Fig. 6, the 1-dimensional linear fitting for a selected peak is the traditional approach for the estimation of the unknown bulk index. In our previous work, we demonstrated that this linear response enables effective modeling through linear or ridge regression when multiple peak shifts are used as inputs, resulting in substantial accuracy improvements over single-feature methods. However, while these isolated peaks behave linearly with respect to index, the full spectrum, comprising many sharp, nonlinear resonances, does not. This fundamental mismatch between spectral shape and linear model assumptions limits the benefit of applying full-spectrum linear regression in the Si-based cases. As we will observe in later sections, unlike our prior study where linear models could leverage the additive effects of multiple linearly shifting features, the current full-spectrum approach yields modest enhancements in silicon structures due to the global nonlinearity embedded in their resonant response.

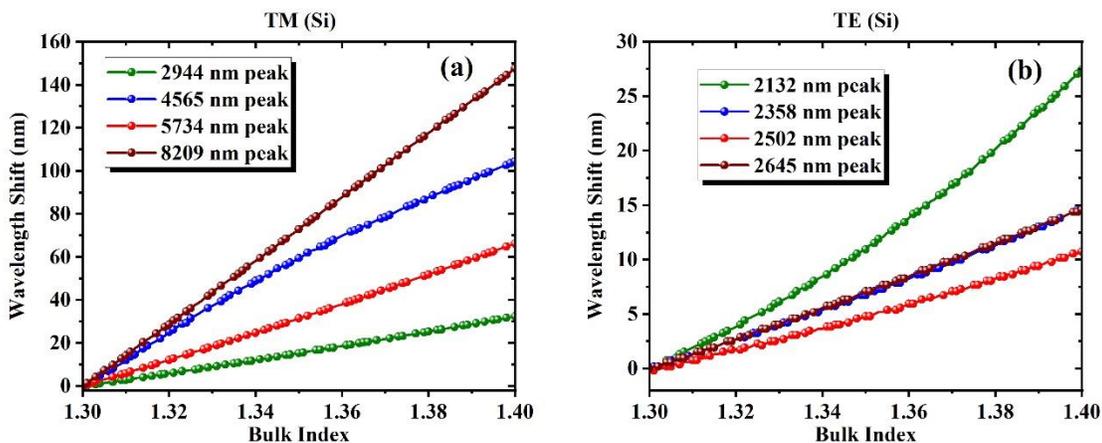

Figure 6. Peak location shift vs bulk index in Si-based meta-grating under (a) TM, and (b) TE polarizarions. Linear behavior is evident while the 8209 nm peak in (a) demonstrates the most linear behavior and the least MSE. Our prior work showed that combining the peaks results in accuracy improvement up to 3 orders of magnitude.

Table 1 summarizes the one-dimensional linear regression performance for selected peaks shown in Figure 6, based on their index-dependent shifts. For each peak, the MSE and MSE standard deviation across five-fold cross-validation are reported. The peaks are chosen from both TM and TE polarization modes of the Si-based structure. Notably, the 8209 nm peak under TM polarization demonstrates the lowest MSE (0.0294 ± 0.0068), indicating a highly linear relationship with refractive index (see Fig. 6(a)). In contrast, some TE-polarized peaks, particularly at shorter wavelengths, exhibit higher MSE values, highlighting their limited utility for accurate linear modeling. These results reinforce the importance of feature selection when using single-variable models and provide a benchmark for evaluating data-driven multivariate approaches.

Table 1. Linear regression 1-dimensional fitting metrics for the peak shifts in Fig. 6.

| TM_Si | | | TE_Si | | |
|---|---|---|---|---|---|
| **Peak Location (nm)[1]** | MSE | | **Peak Location (nm)** | MSE | |
| | Mean | std | | Mean | std |
| **2944** | 0.79198 | 0.16526 | **2132** | 10.035007 | 1.84389 |
| **4565** | 5.89753 | 1.07706 | **2358** | 1.822854 | 0.59077 |
| **5734** | 0.75589 | 0.10943 | **2502** | 2.907157 | 0.63826 |
| **8209** | 0.02944 | 0.00683 | **2645** | 0.562897 | 0.07325 |

Note 1: The peak locations (in nm) mentioned in the table refer to their locations at the bulk index of 1.3.

## 4. Machine learning based full spectrum modeling for precision enhancement

Figure 7 summarizes the modeling performance across the four datasets by comparing (a) the mean squared error (MSE) of the full-spectrum linear regression model using 80 principal components, (b) the corresponding fold enhancement in MSE relative to the best-performing single-feature linear fit, and (c) the MSE values for those best single predictors. These comparisons are made in light of the distinct spectral profiles observed in earlier figures.

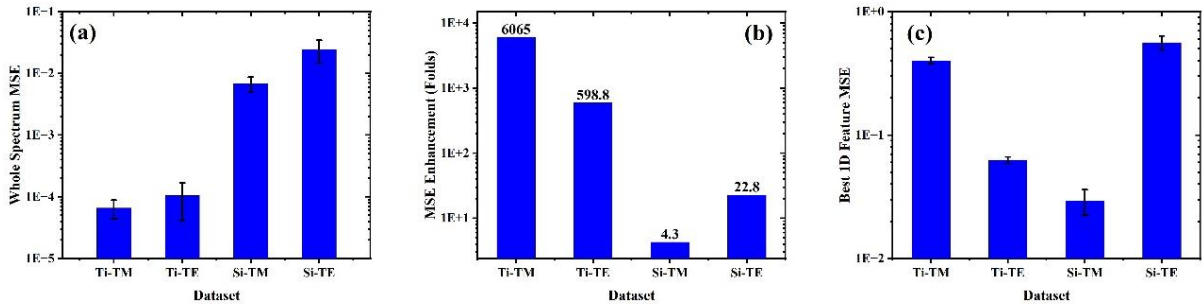

Figure 7. Modeling performance across datasets: (a) Full-spectrum MSE using 80 PCA components, (b) fold improvement over best single-feature fit, and (c) MSE of best single-feature model (peak shifts for Si). Error bars indicate standard deviation from five-fold cross-validation.

In Figure 7(a), Ti-based structures (Ti–TM and Ti–TE) exhibit extremely low MSEs of $6.32 \times 10^{-5}$ and $1.05 \times 10^{-4}$, respectively, when using full-spectrum modeling with 80 PCA components. These values are orders of magnitude lower than those of the Si-based structures: 0.0068 for Si–TM and 0.0247 for Si–TE. This performance discrepancy directly reflects the underlying spectral behavior that Ti spectra exhibit smooth, broadband intensity variations with minimal resonant structure, allowing PCA-based linear regression to effectively capture the index-dependent variation. In contrast, Si spectra are dominated by sharp Mie-type resonances that make the intensity shift nonlinearly with index, making them less compatible with global linear modeling.

Figure 7(b) illustrates the fold enhancement in MSE obtained by transitioning from the best single-feature linear model to full-spectrum PCA-based regression. The Ti–TM dataset shows a 6065-fold reduction in error, followed by Ti–TE with a 598.8-fold improvement. For the Si–TE and Si–TM datasets, the enhancements are more modest (22.8 and 4.3-fold, respectively), reflecting the fact that much of the predictive information in resonance-dominated spectra is already concentrated in a small number of well-aligned peak features.

Figure 7(c) shows the baseline MSEs of the best individual peak predictors: 0.4017 for Ti–TM, 0.0628 for Ti–TE, 0.0294 for Si–TM, and 0.5629 for Si–TE. Notably, while Si–TM yields the lowest single-feature error among the four, its full-spectrum model improves only marginally. This suggests that, despite spectral richness, the nonlinear behavior of its intensity variation limits the effectiveness of linear PCA-based modeling. In contrast, Ti-based datasets benefit substantially from this approach due to their linear and distributed spectral response.

To validate the use of linear regression, a comparison was performed using multiple machine learning algorithms, including Lasso, support vector regression, random forest, and gradient boosting, on the Ti–TE dataset. Linear regression achieved the lowest MSE and was therefore selected for all four datasets to ensure consistency and interpretability. Although the comparison was limited to a representative dataset, the spectral behavior and resulting trends across configurations confirm the appropriateness of the chosen model.

Figure 8 presents the residual plots of the full-spectrum principal component regression (PCR) models across all four datasets, plotted as the difference between the predicted and true target variables against the true target values (index samples from 1.3 to 1.4 with 0.01 steps mapped onto 1 to 101 with 1 seps). These visualizations offer a direct interpretation of prediction error distributions and provide further insight into the compatibility between spectral behavior and linear modeling.

Figures 8(a) and (b), corresponding to Ti–TM and Ti–TE datasets, respectively, reveal extremely tight clustering of residuals around zero, with minimal dispersion and no noticeable bias or systematic trend across the index range. This strongly supports the earlier findings of exceptionally low MSE in these configurations and reinforces that broadband, smooth intensity modulations in Ti-based spectra align well with linear PCA modeling. The residuals appear nearly uniform and homoscedastic, indicating high model fidelity throughout the entire sensing range.

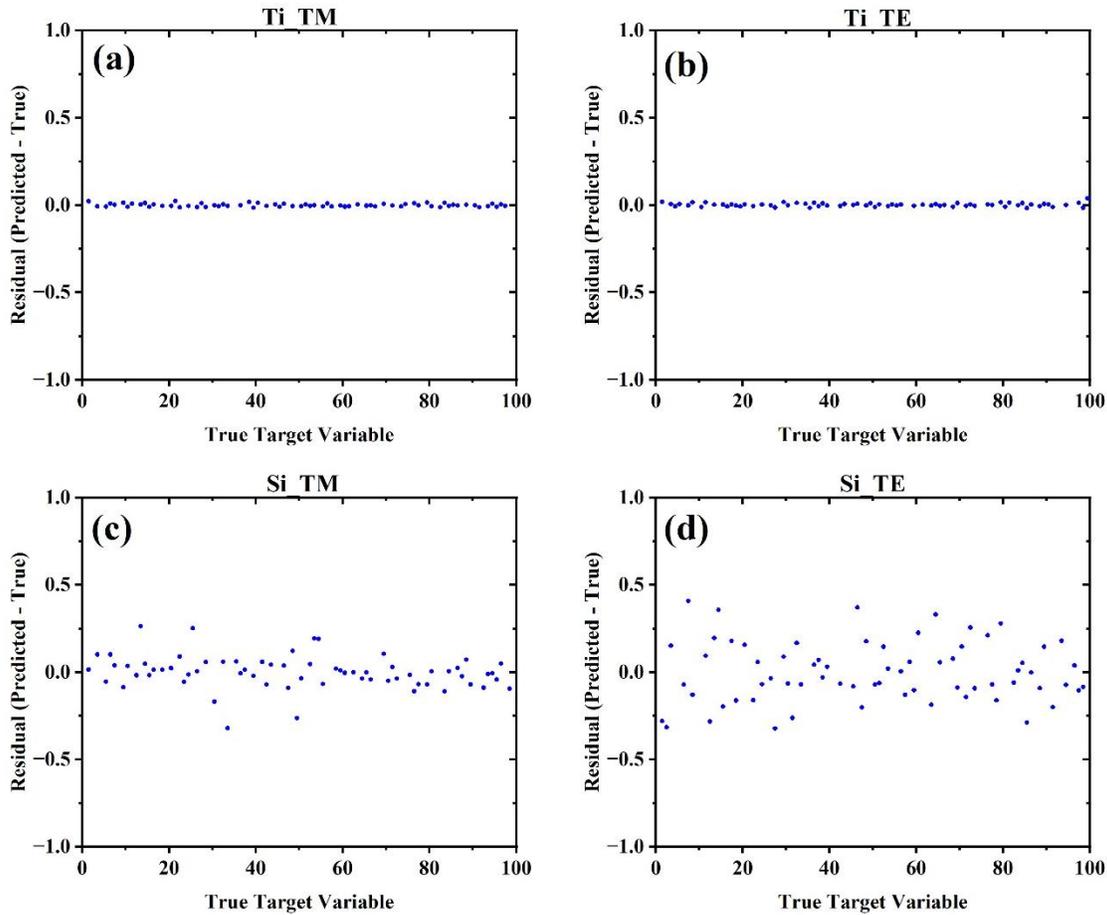

Figure 8. Residual plots from five-fold cross-validated full-spectrum linear regression. Titanium-based datasets (a, b) show minimal error, while silicon-based cases (c, d) exhibit larger, more scattered residuals.

In contrast, Figures 8(c) and (d), associated with Si–TM and Si–TE datasets, show broader distributions of residuals, with Si–TE in particular displaying a visibly higher variance. This pattern is consistent with the elevated MSEs observed in Figure 7(a) and is attributed to the nonlinear intensity-shift behavior inherent to Mie-resonant metasurfaces. The increased variability reflects the challenge of approximating such nonlinearity using a limited number of principal components and a linear regression framework.

Together, these residual plots validate the earlier conclusion: the efficacy of linear modeling is heavily influenced by the underlying spectral transformation mechanism. For Ti metasurfaces with gradual, linear intensity shifts, PCR with 80 components offers precise, unbiased predictions. For Si metasurfaces, while PCA-based modeling still performs reasonably well, nonlinear modeling techniques may be required for further enhancement.

## 5. Conclusion

This study investigated the performance of full spectrum machine learning models, specifically principal component analysis followed by linear regression, in predicting refractive index changes from simulated absorption spectra of metasurface-based biosensors. By comparing four datasets representing different material (Si, Ti) and polarization (TE, TM) configurations, we demonstrated that the predictive accuracy of linear models is strongly influenced by the nature of the underlying spectral changes. Ti-based meta-gratings, characterized by broadband intensity modulation with minimal spectral peak shifts, yielded exceptionally low mean squared errors (MSEs) and tightly clustered residuals. In contrast, Si-based structures with sharp nonlinear Mie resonances showed limited enhancement in performance when modeled using the same linear framework.

The results emphasize the importance of spectral shape in determining the compatibility of linear models with optical sensing data. While PCA-based regression is highly effective for intensity-modulated spectra, its performance is comparatively limited for spectra dominated by resonance shifts. Additionally, the large fold enhancements, exceeding 6000 times for the Ti-based structure and TM polarization, underscore the value of utilizing the entire spectral information rather than relying solely on isolated peak positions. These findings support the use of data-driven feature selection, even in traditional one-dimensional fitting approaches, and suggest that future work incorporating nonlinear modeling could further improve the interpretability and accuracy for complex resonance-based systems.

**Acknowledgements**

The authors thank the support from National Science Foundation under 2225568.

# References

1　Fan, X. *et al.* Sensitive optical biosensors for unlabeled targets: a review. *Anal Chim Acta* **620**, 8-26 (2008). https://doi.org/10.1016/j.aca.2008.05.022

2　Akkilic, N., Geschwindner, S. & Hook, F. Single-molecule biosensors: Recent advances and applications. *Biosens Bioelectron* **151**, 111944 (2020). https://doi.org/10.1016/j.bios.2019.111944

3　Chen, C. & Wang, J. Optical biosensors: an exhaustive and comprehensive review. *Analyst* **145**, 1605-1628 (2020). https://doi.org/10.1039/c9an01998g

4　Haque, E., Anwar Hossain, M., Namihira, Y. & Ahmed, F. Microchannel-based plasmonic refractive index sensor for low refractive index detection. *Appl Opt* **58**, 1547-1554 (2019). https://doi.org/10.1364/AO.58.001547

5　Homola, J. Present and future of surface plasmon resonance biosensors. *Anal Bioanal Chem* **377**, 528-539 (2003). https://doi.org/10.1007/s00216-003-2101-0

6　Jagerska, J., Zhang, H., Diao, Z., Le Thomas, N. & Houdre, R. Refractive index sensing with an air-slot photonic crystal nanocavity. *Opt Lett* **35**, 2523-2525 (2010). https://doi.org/10.1364/OL.35.002523

7　Jaksic, Z., Vukovic, S., Matovic, J. & Tanaskovic, D. Negative Refractive Index Metasurfaces for Enhanced Biosensing. *Materials (Basel)* **4**, 1-36 (2010). https://doi.org/10.3390/ma4010001

8　Kaur, B., Kumar, S. & Kaushik, B. K. Recent advancements in optical biosensors for cancer detection. *Biosens Bioelectron* **197**, 113805 (2022). https://doi.org/10.1016/j.bios.2021.113805

9　Konopsky, V. N. & Alieva, E. V. A biosensor based on photonic crystal surface waves with an independent registration of the liquid refractive index. *Biosens Bioelectron* **25**, 1212-1216 (2010). https://doi.org/10.1016/j.bios.2009.09.011

10　Lavin, A. *et al.* On the Determination of Uncertainty and Limit of Detection in Label-Free Biosensors. *Sensors (Basel)* **18** (2018). https://doi.org/10.3390/s18072038

11　Liu, Y., Zhou, W. & Sun, Y. Optical Refractive Index Sensing Based on High-Q Bound States in the Continuum in Free-Space Coupled Photonic Crystal Slabs. *Sensors (Basel)* **17** (2017). https://doi.org/10.3390/s17081861

12      Lodewijks, K., Van Roy, W., Borghs, G., Lagae, L. & Van Dorpe, P. Boosting the figure-of-merit of LSPR-based refractive index sensing by phase-sensitive measurements. *Nano Lett* **12**, 1655-1659 (2012). https://doi.org/10.1021/nl300044a

13      Maksimov, D. N., Gerasimov, V. S., Romano, S. & Polyutov, S. P. Refractive index sensing with optical bound states in the continuum. *Opt Express* **28**, 38907-38916 (2020). https://doi.org/10.1364/OE.411749

14      Perera, C. *et al.* Highly compact refractive index sensor based on stripe waveguides for lab-on-a-chip sensing applications. *Beilstein J Nanotechnol* **7**, 751-757 (2016). https://doi.org/10.3762/bjnano.7.66

15      Shangguan, Q. *et al.* Design of Ultra-Narrow Band Graphene Refractive Index Sensor. *Sensors (Basel)* **22** (2022). https://doi.org/10.3390/s22176483

16      Wang, Q. *et al.* Research advances on surface plasmon resonance biosensors. *Nanoscale* **14**, 564-591 (2022). https://doi.org/10.1039/d1nr05400g

17      White, I. M. & Fan, X. On the performance quantification of resonant refractive index sensors. *Opt Express* **16**, 1020-1028 (2008). https://doi.org/10.1364/oe.16.001020

18      Yan, Z. *et al.* Multi-Structure-Based Refractive Index Sensor and Its Application in Temperature Sensing. *Sensors (Basel)* **25** (2025). https://doi.org/10.3390/s25020412

19      Banerjee, A., Maity, S. & Mastrangelo, C. H. Nanostructures for Biosensing, with a Brief Overview on Cancer Detection, IoT, and the Role of Machine Learning in Smart Biosensors. *Sensors (Basel)* **21** (2021). https://doi.org/10.3390/s21041253

20      Hassan, M. M. *et al.* Progress of machine learning-based biosensors for the monitoring of food safety: A review. *Biosens Bioelectron* **267**, 116782 (2025). https://doi.org/10.1016/j.bios.2024.116782

21      Khanal, B., Pokhrel, P., Khanal, B. & Giri, B. Machine-Learning-Assisted Analysis of Colorimetric Assays on Paper Analytical Devices. *ACS Omega* **6**, 33837-33845 (2021). https://doi.org/10.1021/acsomega.1c05086

22      Kokabi, M., Tahir, M. N., Singh, D. & Javanmard, M. Advancing Healthcare: Synergizing Biosensors and Machine Learning for Early Cancer Diagnosis. *Biosensors (Basel)* **13** (2023). https://doi.org/10.3390/bios13090884

23   Pan, X. *et al.* Machine Learning-Assisted High-Throughput Identification and Quantification of Protein Biomarkers with Printed Heterochains. *J Am Chem Soc* **146**, 19239-19248 (2024). https://doi.org/10.1021/jacs.4c04460

24   Schackart, K. E., 3rd & Yoon, J. Y. Machine Learning Enhances the Performance of Bioreceptor-Free Biosensors. *Sensors (Basel)* **21** (2021). https://doi.org/10.3390/s21165519

25   Zhang, K. *et al.* Machine Learning-Reinforced Noninvasive Biosensors for Healthcare. *Adv Healthc Mater* **10**, e2100734 (2021). https://doi.org/10.1002/adhm.202100734

26   Ajala, S., Muraleedharan Jalajamony, H., Nair, M., Marimuthu, P. & Fernandez, R. E. Comparing machine learning and deep learning regression frameworks for accurate prediction of dielectrophoretic force. *Sci Rep* **12**, 11971 (2022). https://doi.org/10.1038/s41598-022-16114-5

27   Aalizadeh, M., Azmoudeh Afshar, M. & Fan, X. Machine Learning Enabled Multidimensional Data Utilization Through Multi-Resonance Architecture: A Pathway to Enhanced Accuracy in Biosensing. *arXiv* **2412** (2024). https://doi.org/https://arxiv.org/abs/2412.20245

28   Richter, F. U. *et al.* Gradient High-Q Dielectric Metasurfaces for Broadband Sensing and Control of Vibrational Light-Matter Coupling. *Adv Mater* **36**, e2314279 (2024). https://doi.org/10.1002/adma.202314279

29   Rodrigo, D. *et al.* Resolving molecule-specific information in dynamic lipid membrane processes with multi-resonant infrared metasurfaces. *Nat Commun* **9**, 2160 (2018). https://doi.org/10.1038/s41467-018-04594-x

30   Shen, Z. & Du, M. High-performance refractive index sensing system based on multiple Fano resonances in polarization-insensitive metasurface with nanorings. *Opt Express* **29**, 28287-28296 (2021). https://doi.org/10.1364/OE.434059

31   Tittl, A., John-Herpin, A., Leitis, A., Arvelo, E. R. & Altug, H. Metasurface-Based Molecular Biosensing Aided by Artificial Intelligence. *Angew Chem Int Ed Engl* **58**, 14810-14822 (2019). https://doi.org/10.1002/anie.201901443

32   Zhang, C. *et al.* Terahertz toroidal metasurface biosensor for sensitive distinction of lung cancer cells. *Nanophotonics* **11**, 101-109 (2022). https://doi.org/10.1515/nanoph-2021-0520

33  Zhang, Z. *et al.* The Antibody-Free Recognition of Cancer Cells Using Plasmonic Biosensor Platforms with the Anisotropic Resonant Metasurfaces. *ACS Appl Mater Interfaces* **12**, 11388-11396 (2020). https://doi.org/10.1021/acsami.0c00095

34  Aalizadeh, M., Serebryannikov, A. E., Ozbay, E. & Vandenbosch, G. A. E. A simple Mie-resonator based meta-array with diverse deflection scenarios enabling multifunctional operation at near-infrared. *Nanophotonics* **9**, 4589-4600 (2020). https://doi.org/doi:10.1515/nanoph-2020-0386